\documentclass[twocolumn]{webofc}
\usepackage[varg]{txfonts}
\usepackage{graphicx}
\usepackage{subfigure}
\usepackage{amsmath}
\usepackage{subfigure}
\usepackage{natbib}

\begin{document}
\title{Strength and energy consumption of inherently anisotropic rocks at failure}

\author{\firstname{David} \lastname{Cantor}\inst{1,2}\fnsep\thanks{\email{david.cantor@polymtl.ca}} 
        \and
        \firstname{Carlos} \lastname{Ovalle}\inst{1,2}
        \and
        \firstname{Emilien} \lastname{Az\'{e}ma}\inst{3,4}
        % etc.
}

\institute{Department of Civil, Geological and Mining Engineering, Polytechnique Montr\'{e}al, Qu\'{e}bec, Canada
\and
Research Institute of Mining and Environment, RIME UQAT-Polytechnique, Montr\'{e}al, Qu\'{e}bec, Canada
\and
LMGC, Universit\'{e} de Montpellier, CNRS, Montpellier, France
\and
Institut Universitaire de France (IUF), Paris, France
}

\abstract{
Using a discrete-element approach and a bonding interaction law, we model and test crushable inherently anisotropic structures reminiscent of the layering found in sedimentary and metamorphic rocks. 
By systematically modifying the level of inherent anisotropy, we characterize the evolution of the failure strength of circular rock samples discretized using a modified Voronoi tesselation under diametral point loading at different orientations relative to the sample's layers. 
We characterize the failure strength, which can dramatically increase as the loading becomes orthogonal to the rock layers. 
We also describe the evolution of the macroscopic failure modes as a function of the loading orientation and the energy consumption at fissuring. 
Our simulation strategy let us conclude that the length of bonds between Voronoi cells controls the energy being consumed in fissuring the rock sample, although failure modes and strength are considerably changing. 
We end up this work showing that the microstructure is largely affected by the level of inherent anisotropy and loading orientation. 
}
\maketitle
\section{Introduction}\label{intro}
The failure strength of rocks is linked to the genesis and mineral composition and organization in space. 
In sedimentary or metamorphic rocks, the layering of minerals strongly affects  joint family characteristics, bonds, and fissuring (i.e., an inherent geometrical anisotropy is imprinted in these materials by formation, tear and wear). 
Ultimately, the mechanical behavior of anisotropic rocks may strongly vary due to the layering properties and the loading orientation on rock samples \cite{Hoek1964,Amadei1996,Karakul2010,Khanlari2015,Pouragha2020}. 
Typically, the failure strength of rock cores presenting an anisotropic inner structure is tested under diametral point load, also known as the \emph{brazilian test} in which the loading orientation is varied from totally aligned to the layering up to orthogonal. 
In general, the failure strength turns out to be maximal for the orthogonal loading and minimal for diagonal-like load orientation relative to the layering. 

More recently, numerical methods, mostly based on the discrete-element approach, have been employed to model and analyze the behavior of anisotropic rock given its capabilities of reproducing failure and crushing (see, for instance, Refs. \cite{Pouragha2020, Xu2020}). 
These methods have added valuable information concerning the modeling and failure mechanism of anisotropic brittle structures. 
Nonetheless, the microstructure characteristics, failure modes evolution, and energy consumption partitions by rupture modes, still lack an extended analysis. 
This is not a trivial modeling and analysis given that anisotropic rock structures need a numerical approach that allows samples to break in multiple irregular pieces of diverse shape and size, while measuring energy consumption as fissures appears within the bulk. 
As we will show in this paper, this is possible by using a discrete-element approach and a modified Voronoi decomposition of samples, allowing us to create cells reminiscent of the mineral organization, and letting us finely control the level of inherent anisotropy. 
The outcomes of this research axis are large and may cover large scales for tunnel and other underground earthworks, up to the failure of granulates and conglomerates for powder technology.

This paper is organized as follows. 
In Sec. \ref{Numerics}, we describe the numerical method employed to build and test anisotropic rock samples. 
In Sec. \ref{Macro}, we present the characterization of the failure strength as a function of the level of inherent anisotropy and loading orientation. 
We then describe the failure modes evolution as a function of the loading orientation in Sec. \ref{Modes}.
We also quantify the energy consumption at fissuring from the beginning of the loading up to the breakage in Sec. \ref{Energy}. 
Finally, in Sec. \ref{Connectivity}, we show how the loading affects the microstructure by comparing the intact connectivity of cells and that on the onset of failure. 
We finish this work with some conclusions and perspectives. 

\section{Numerical model}\label{Numerics}
We use the discrete-element method known as contact dynamics (CD) \cite{Jean1999,Dubois2018} to simulate breakable brittle rock cores.
The CD method is a discrete approach in which rigid bodies interact with solid contact, friction, and cohesion. 
In particular, the CD method is an implicit method in which the equations of motion are applied to a collection of bodies in contact for which velocities and interaction forces are computed on a time-stepping scheme. 

\begin{figure}
  \centering
  \includegraphics[width=0.8\linewidth]{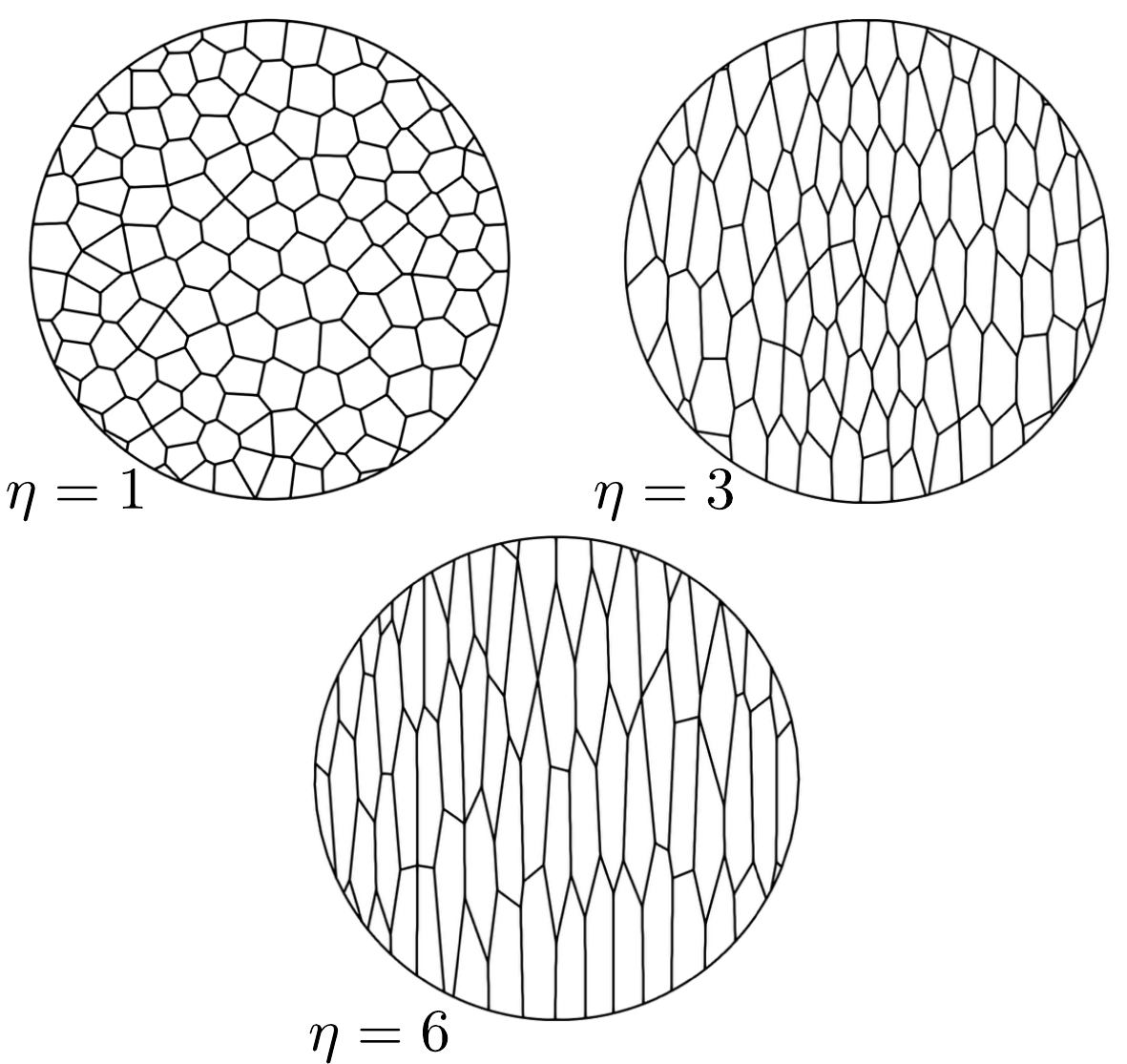}
  \caption{Screenshots of samples presenting a different cell aspect ratio $\eta$.}
  \label{fig:samples}
\end{figure}

In order to build anisotropic structures, we use a Voronoi tessellation on circular samples, so the resulting cells' geometry translates the inherent anisotropy level of a rock. 
Given that a random Voronoi tessellation produces cells of varied shapes and sizes, we modify the tessellation procedure to finely control the level of anisotropy. 
First, we employ a centroidal Voronoi tessellation in which iteratively we reuse the cells' centroid as seeding for next meshings. 
By doing so, we can obtain cells of similar geometries. 
Then, we elongate all the cells in a given direction so we can compute their average shape ratio $\eta=h/L$, with $h$ and $L$ the average small and larger dimensions.
We produce a set of samples in which $\eta$ varied from one to six in steps of 1 (see Fig. \ref{fig:samples}).

To provide mechanical strength to the assembly of cells, we used the bonded-cell method approach previously used in the simulation and analysis of crushable grains \cite{Nguyen2015,Cantor2016,Huillca2020}. 
By adding a bonding law at the interfaces between cells (note that these interfaces are exclusively edge-edge contacts), the cells' interfaces can resist tensile and shearing stresses. 
This needs the definition of two values of cohesion $C_n$ and $C_t$, that, for the sake of simplicity, we fix to the same value of $10$ kPa. 
Additionally, we preset a debonding distance $\delta_c$ necessary to break a cell-cell bond. 
In other words, a fissure between two cells can only be broken if once the stress threshold $C_n$ or $C_t$ is reached, they present a gap or relative sliding equals to $\delta_c$. 
The introduction of this debonding distance is not arbitrary. 
It allows us to represent the progressive debonding of two cells, and to characterize the type of material we are analyzing. 
Following fracture mechanics theory, we then choose a typical value of surface energy for silicate materials $\gamma=50 \mathrm{ J/m}^2$ \cite{Jones2019}, and then the debonding distance is simply defined as $\delta_c= 2 \gamma/C_n$. 

Note that this construction strategy can generate configurations with varying number of cells depending on the anisotropy level. 
Several studies have used the cells' strategy to reproduce fragmentation and have spotlighted the role of the number of cells on the mechanical response. 
As a general observation, adding cells reduces the final strength of the samples, which is in agreement with the fact that more interactions between cells adds more failure potential paths. 
Nonetheless, to make the different samples comparable, we set the total length of cohesive interactions to be constant. 

To test these samples, we used the diametral point load in which two rigid platens gradually increasing the applied force. 
We also varied the orientation of the load $\theta$ respect to the layering of the cells, as shown in the inset of Fig. \ref{fig:strength}
Finally, we perform our simulations by including the bonded-cell method in the simulation platform LMGC90, developed at the Universit\'e de Montpellier \cite{LMGC90Web,Dubois2011}. 

\section{Failure strength}\label{Macro}
\begin{figure}
  \centering
  \includegraphics[width=0.92\linewidth]{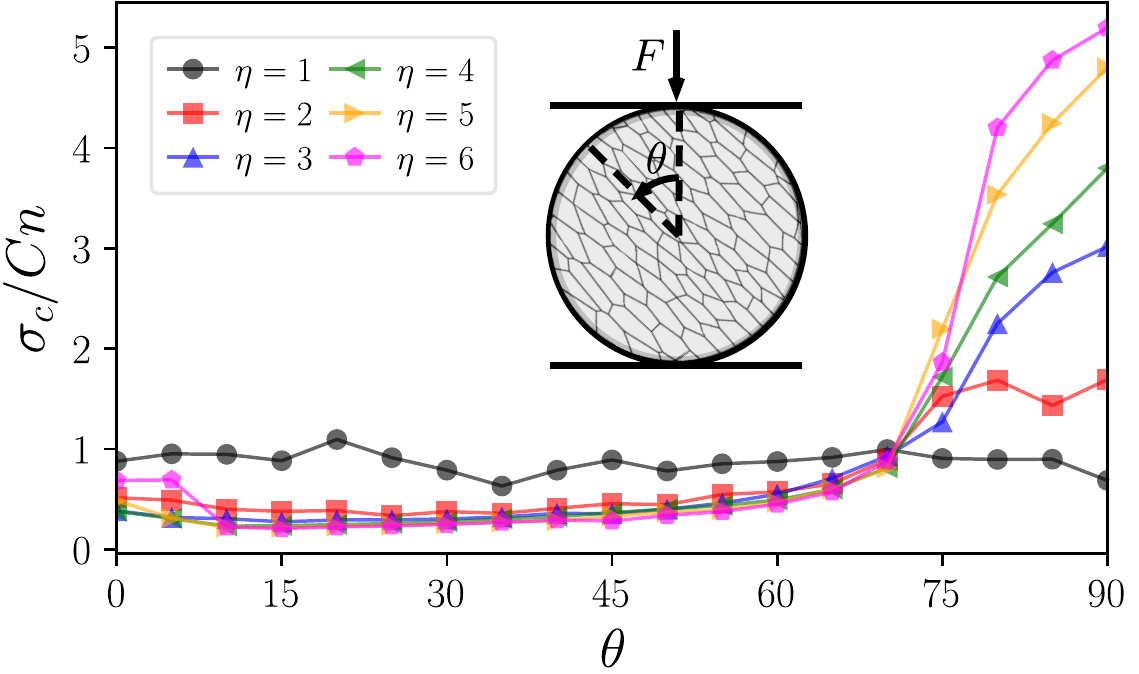}
  \caption{Failure strength as a function of the level of inherent anisotropy $\eta$ and the loading orientation $\theta$. In the inset: schematic representation of our loading procedure and the relative loading orientation relative to the rock layering. }
  \label{fig:strength}
\end{figure}

The failure strength of the samples is found as the loading reaches a critical force value $F_c$, for which the structure can no longer bear load and collapses. 
By using the tensile failure criterion for circular geometries, we can characterize the failure strength $\sigma_c$ as 
\begin{equation}\label{eq:sigma_c}
  \sigma_c = \frac{2F_c}{\pi d},
\end{equation}
with $d$ the diameter of the sample. 
Figure \ref{fig:strength} gathers the values of crushing strength as a function of the level of inherent anisotropy and loading orientation normalized by the cohesive strength $C_n$. 
We can remark that for a non anisotropic structure $\eta=1$, the strength remains near $C_n$.
Once some degree of anisotropy is introduced in the cells' configuration, the strength substantially varies. 
First, for loading orientations in the range $[0^{\circ},70^{\circ}]$ the strength is lower than $C_n$ and very similar between values of $\eta$. 
For orientations beyond $75^{\circ}$, the failure strength rapidly increases and reaches higher values for larger $\eta$.
This behavior is consistent with laboratory testing as well as  the minimum failure strength for an orientation around $30^{\circ}$ after simple stress considerations \cite{Griffith1921}.

In the following, we explore the mechanisms behind this strong difference in failure strength among samples and loading orientations. 

\section{Failure modes}\label{Modes}
\begin{figure}%[h!]
    \centering
    \subfigure {\label{fig:frac00}(a)
    \includegraphics[width=0.35\columnwidth]{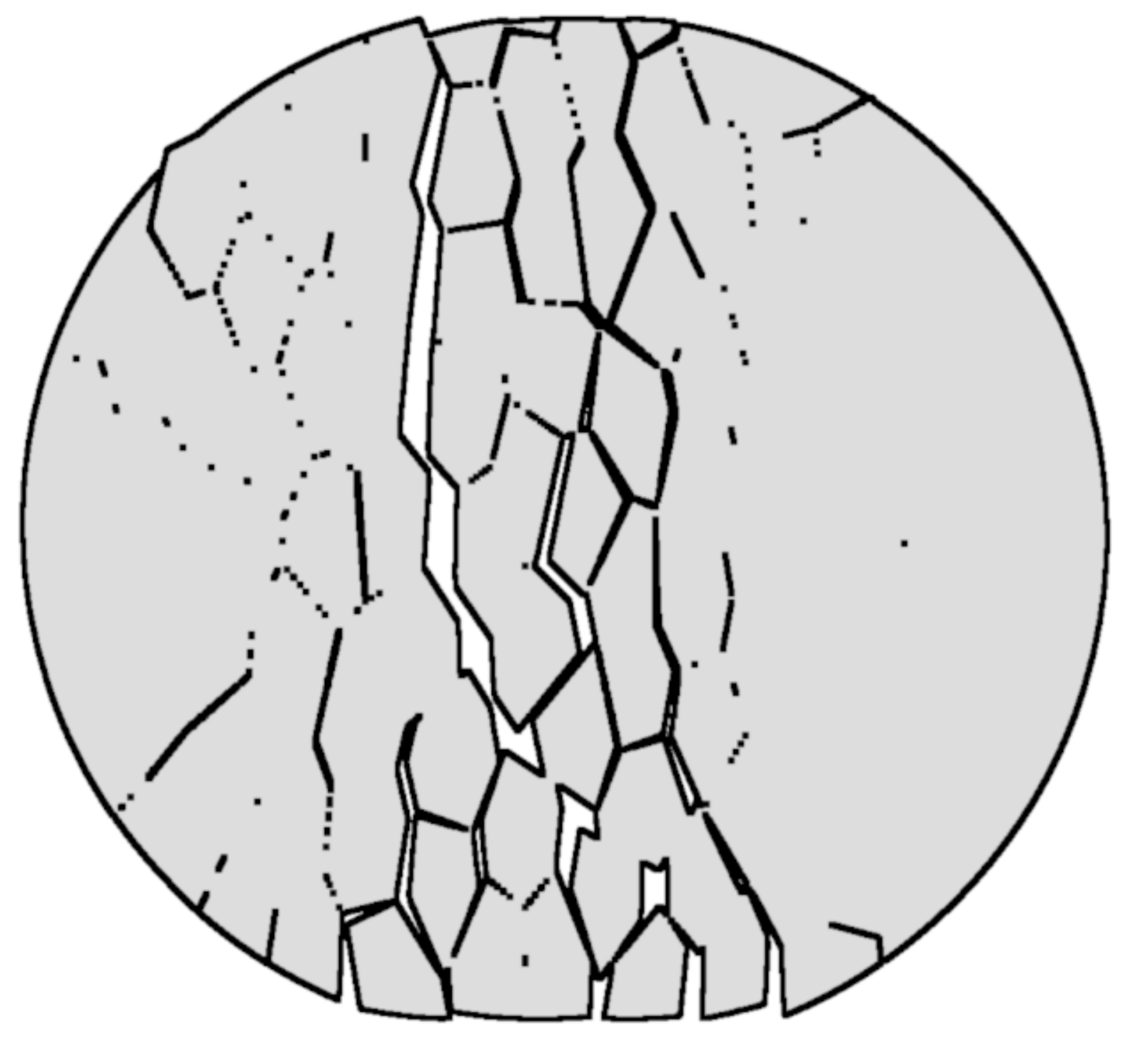}}
    \subfigure {\label{fig:frac30}(b)
    \includegraphics[width=0.35\columnwidth]{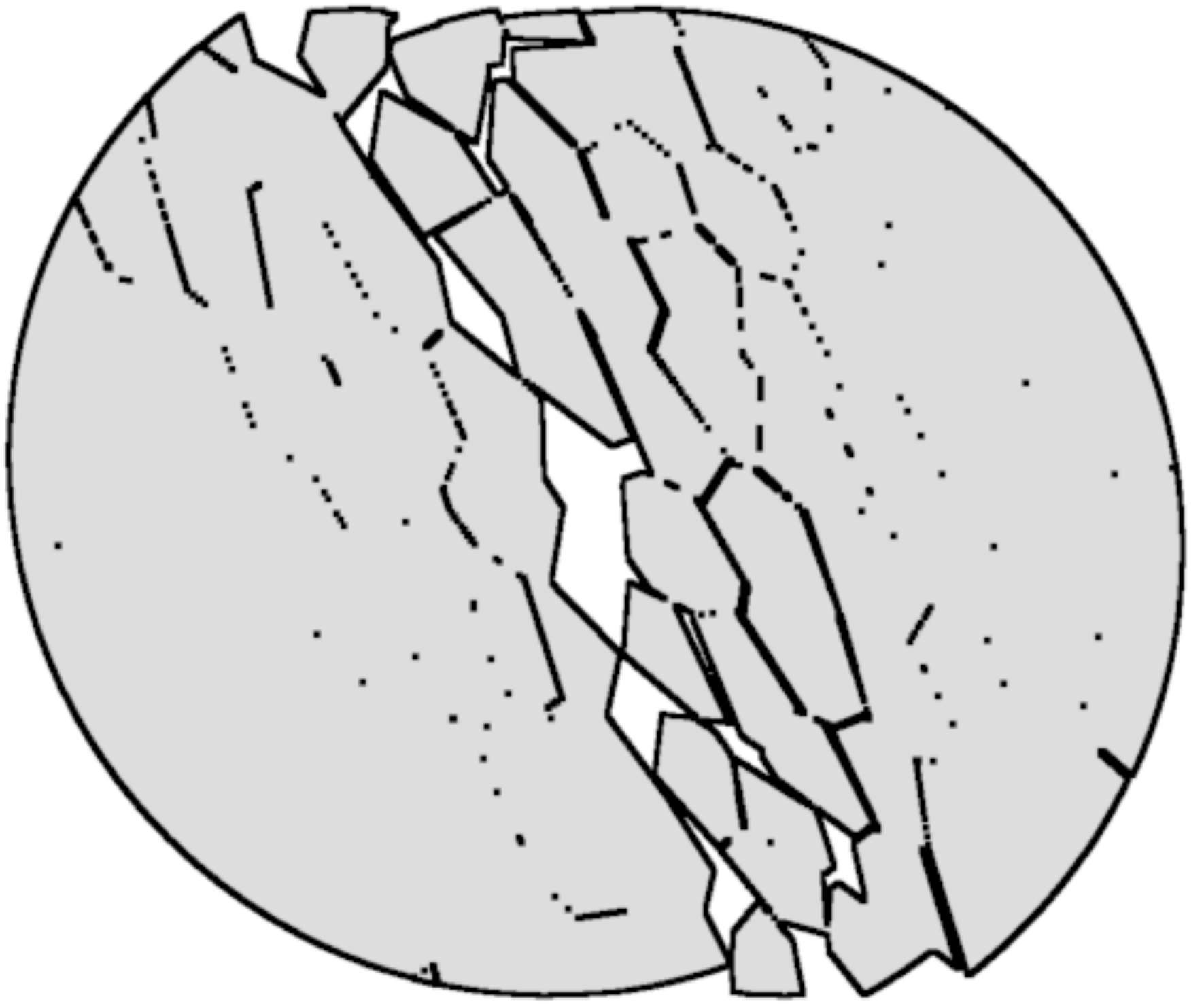}}
    \subfigure {\label{fig:frac60}(c)
    \includegraphics[width=0.35\columnwidth]{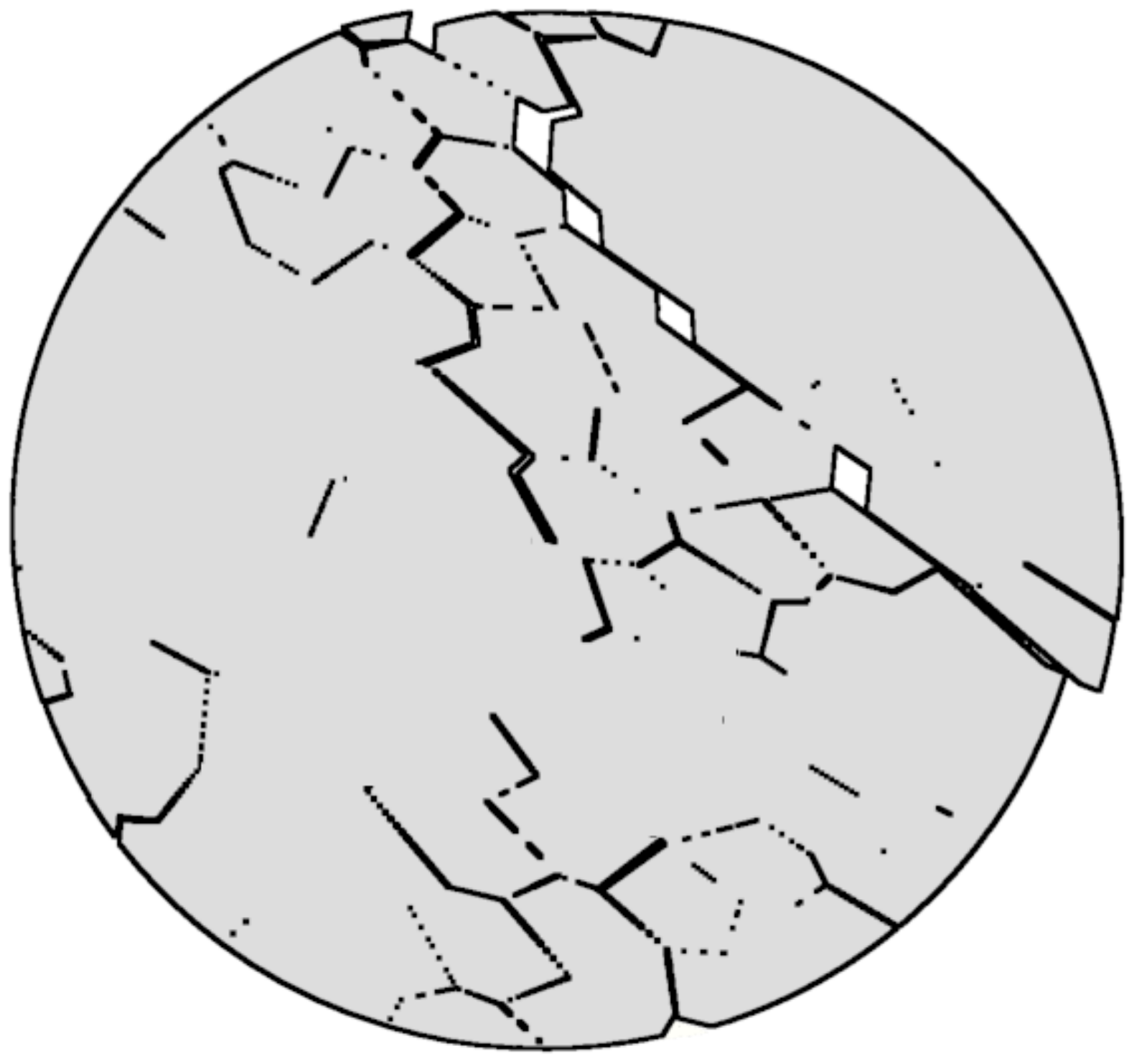}}
    \subfigure {\label{fig:frac90}(d)
    \includegraphics[width=0.35\columnwidth]{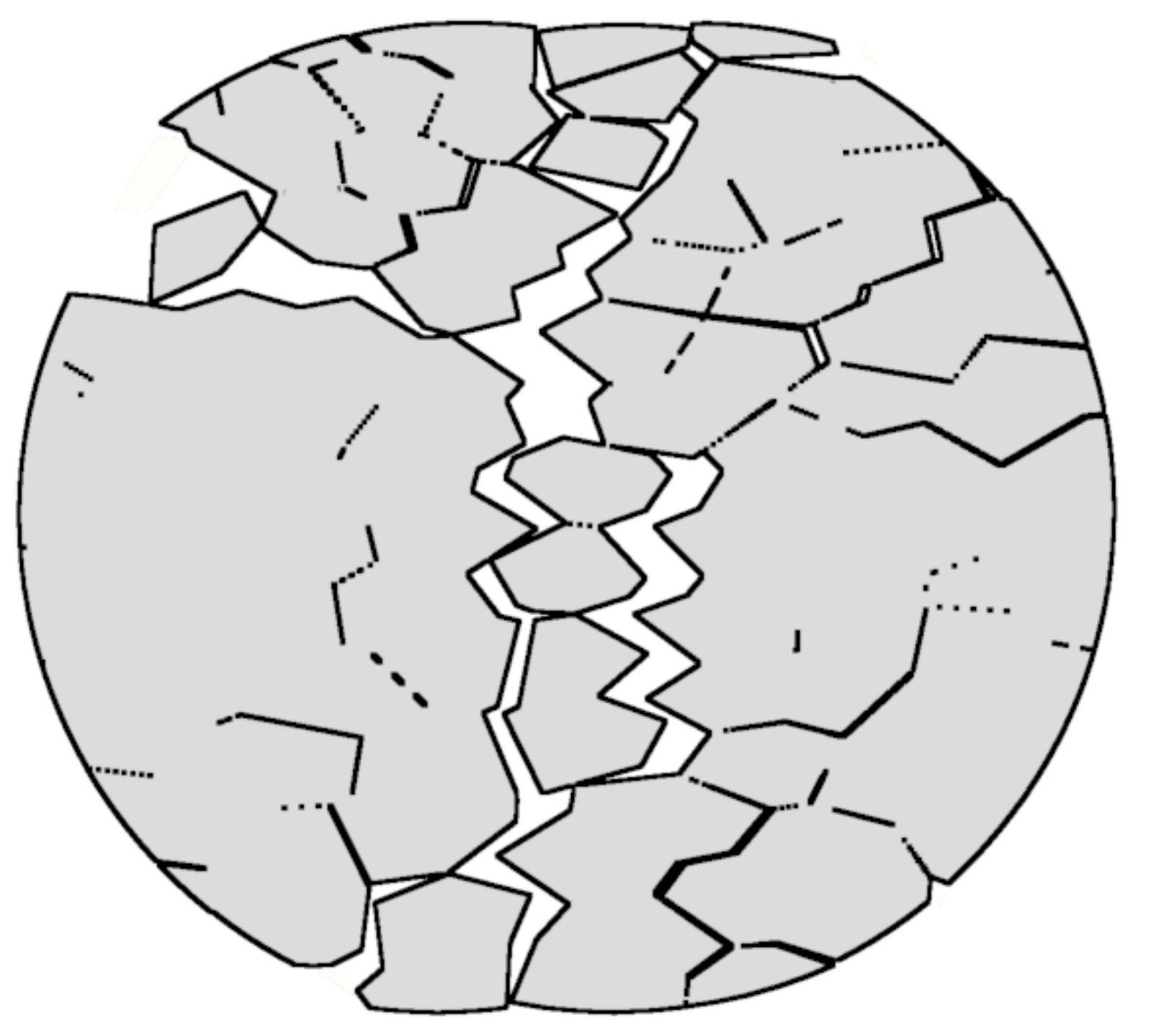}}
    \caption{Evolution of the failure mechanism as the loading orientation increases (a) $\theta = 0^{\circ}$, (b) $\theta = 30^{\circ}$,(c) $\theta = 60^{\circ}$ and (d) $\theta = 90^{\circ}$ for the sample with inherent anisotropy $\eta=2$.}
    \label{fig:failure_modes}
\end{figure}

The first qualitative characteristic we can highlight is the evolution of failure modes as the loading orientation varies. 
Figure \ref{fig:failure_modes} presents screenshots of the samples with $\eta=2$ for increasing $\theta$. 

For a loading orientation matching the cell's elongation (i.e., $\theta = 0^{\circ}$), the fissures are mostly vertical and percolating between the walls. 
For such a case, we can deduce that tensile stresses in the orthogonal direction to the loading were responsible of this failure. 
For loading orientations $\theta = 30^{\circ}$ and $\theta = 60^{\circ}$, the failure paths are different and diagonal respect to the application point or force. 
The relative displacement between cells suggest that shearing is a major mechanism for debonding of cells. 
Finally, for the case $\theta = 90^{\circ}$, we can observe that the fissuring paths are more complex, presenting larger amounts of damage near the contact with the loading plates, nonetheless a vertical series of cracks may suggest that tensile stresses were also important for this failure. 
This observations are in agreement with the drop of failure strength for anisotropic structures presenting diagonal fissuring trajectories. 
Indeed, mobilization of cells end up being traveling shorter distances for such cases. 
Nonetheless, the rapid increase of strength for more orthogonal-like loading conditions is impossible to grasp from the visualizations. 

\section{Energy consumption}\label{Energy}
We can analyze the consumption of energy by partitions being consumed in tensile failures and shear failures. 
The total energy $E_c$ being consumed at fissuring can be split as 
\begin{equation}
  E_c = E_n + E_t = C_n \delta_c \sum_{\forall c^*_n} l_{c^*_n} + C_t \delta_c \sum_{\forall c^*_t} l_{c^*_t},
\end{equation}
with $c^*_n$ and $c^*_t$, the number of bonds broken in traction or shearing, and $l_{c^*_n}$ and $l_{c^*_t}$ their corresponding lengths. 

\begin{figure}
  \centering
  \includegraphics[width=0.92\linewidth]{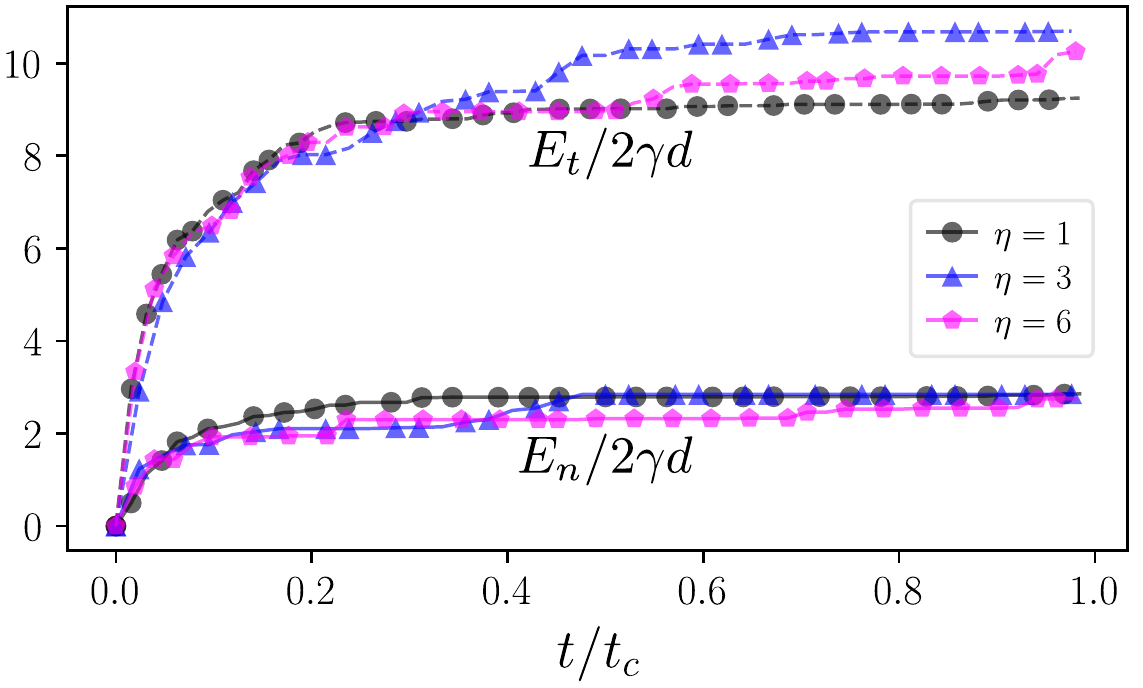}
  \caption{Evolution in time of cumulative energy consumption in mode traction and shearing for some of the samples at $\theta = 0$.}
  \label{fig:EvolE}
\end{figure}

Figure \ref{fig:EvolE} presents the evolution of $E_n$ and $E_t$ during the loading of samples with $\eta = 1,3,$ and $6$, up to the moment of failure ($t_c$) and for loading orientation $\theta = 0$. 
The data is normalized by the nominal energy consumption for a vertical straight failure path (i.e., $2 \gamma d$). 
In effect, the ratio $E_c/2\gamma d$ translates the tortuosity of the failure path in the cell configuration. 
As a matter of fact, fixing the total bonding length between cells to be constant produces failures consuming similar energy amounts independently of $\eta$. 
Note that this is the failure mechanism at the cells' scale and is not necessarily related to the `apparent' macroscopic failure mode of the assembly. 
The trend of these curves also proves that no burst of energy occurs at failure. 
Conversely, the cohesive bonds are progressively broken (i.e., fissuring) up to a critical energy for which the particle collapses. 
This fact is in agreement with the failure mechanics theory and energy criticality at failure.

\begin{figure}
  \centering
  \includegraphics[width=0.92\linewidth]{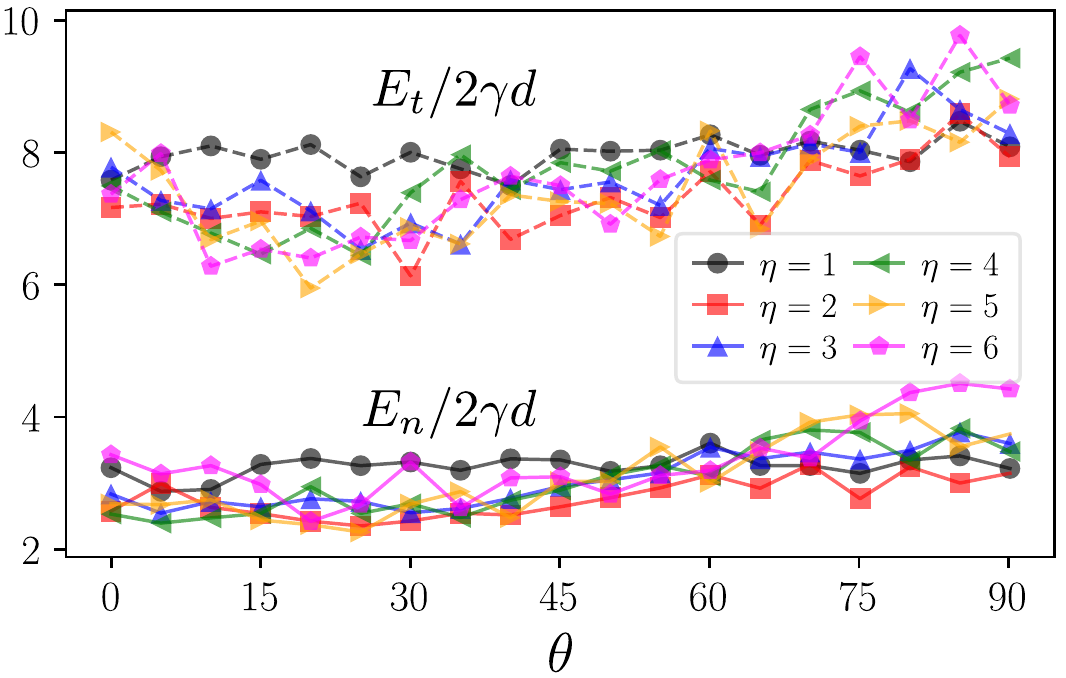}
  \caption{Total energy consume at failure in traction and shearing modes as a function of $\theta$ and different $\eta$.}
  \label{fig:energy_tot}
\end{figure}

In Fig. \ref{fig:energy_tot} we observe the total energy consumed by mode (traction or shearing) as a function of the loading orientation and $\eta$.
We see that, in general, more fissuring occurs in shearing mode than in traction. 
A slight increase of $E_n$ and $E_t$ also seems to occur for increasing loading orientation $\theta$.
Nonetheless, the variation is relatively small compared with the order of magnitude of the energy, and the fact that the ratio between energies remains roughly the same for all the values of $\eta$ and $\theta$.

\section{Cells' connectivity}\label{Connectivity}
Using the coordination number, defined as $Z = 2 N_c/N_{cl}$, with $N_c$ the number of bonds between cells, and $N_{cl}$ the number of cells, we compute the average number of neighboring interactions per cell. 
This parameter makes part of the signature of mineral organization within geomaterials. 

Note that a subtle difference in coordination number can be found between the intact coordination number (i.e., at the beginning of the loading) and the coordination number instants prior failure due to fissuring. 
As we rather characterize the onset of failure as the state bearing $\sigma_c$, let us consider the cohesive bonds on the onset of failure $N^*_c$ as the effective number of interactions, so $Z^* = 2 N_c^*/N_{cl}$. 

\begin{figure}
  \centering
  \includegraphics[width=0.92\linewidth]{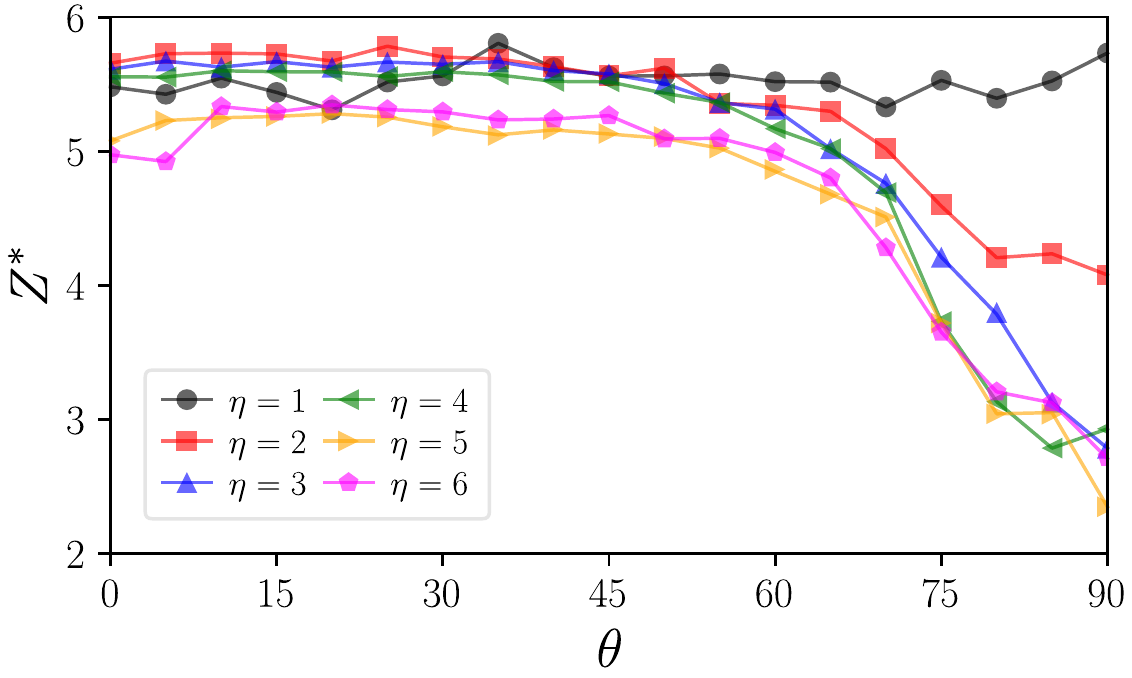}
  \caption{Evolution of the coordination number at the onset of failure as a function of $\eta$ and loading orientation $\theta$.}
  \label{fig:z}
\end{figure}

Figure \ref{fig:z} presents the averaged coordination number on the onset of failure as a function of $\eta$. 
We observe that the connectivity decreases as $\theta$ and $\eta$ grow. 
This behavior is due to the fact that orthogonal loading respect to the cell's main orientation can reach larger forces through column-like structure. 
In the meantime, larger values of $\eta$ imply fewer contacts within the assembly, so the lost of a given number of bonds imply a stronger drop of $Z^*$. 
Note that a centroidal Voronoi tessellation tends to produce hexagonal-like configurations of cells, then a value of $Z \simeq 6$ is expected for intact samples. 
Finally, the important variation of $Z^*$ with $\eta$ and $\theta$ suggests that the number of bonds participating in the load transmission can largely change at the onset of failure, implying strong variation in the microstructural characteristics producing $\sigma_c$ at the macroscopic scale. 

\section{Conclusions}\label{Conclusions}
In the frame of the contact dynamics method and the bonded-cell approach, we built and tested crushable samples recalling inherent anisotropic geometries often found in sedimentary or metamorphic rocks. 
This approach let us define a level of anisometry $\eta$ of the internal structure of circular samples divided in cells via a modified Voronoi tessellation. 
By applying a diametral loading, we were able to characterize the failure strength as a function of $\eta$ and the loading orientation $\theta$. 
Our results, in agreement with experimental observations and fracture mechanics theory, showed that the strength decreases for orientations $\theta<70^{\circ}$ once an inherent anisotropic structure is included in the samples. 
But as $\theta$ gets values beyond $75^{\circ}$, the failure strength rapidly increases with $\eta$. 

Despite these variations of failure strength, our simulations showed that fixing the total length of cohesive bonds between cells determines the amount of energy being consumed at fissuring, which, ultimately, has not relation with the resistance. 
Then, another microstructural elements should be at the origin of the failure strength. 
We explore this microstructure using the coordination number $Z$. 
We observed that this parameter it strongly varies with $\eta$ and $\theta$. 
But, counterintuitively, it seems that stronger inherent anisotropic structures also are those who present lower connectivity of their components. 
Future research should focus on the force transmission mechanisms and should finely quantify the inherent and induced anisotropy characteristics to better asses the sources of the macroscopic observations from the cells and bonds scale. \\

\small{This research benefited from the financial support of Fonds de recherche du Québec - Nature et technologies [Ref. 2020-MN-281267], and the industrial partners of RIME UQAT-Polytechnique (irme.ca/en).}

\normalsize

\bibliography{biblio}

\begin{thebibliography}{15}

\bibitem{Hoek1964}
E.~Hoek, J. S. Afr. I. Min. Metall. \textbf{64}, 501 (1964)

\bibitem{Amadei1996}
B.~Amadei, Int. J. Rock Mech. Min. Sci \textbf{33}, 293 (1996)

\bibitem{Karakul2010}
H.~Karakul, R.~Ulusay, N.S. Isik, Int. J. Rock Mech. Min. Sci. \textbf{47}, 657
  (2010)

\bibitem{Khanlari2015}
G.~Khanlari, B.~Rafiei, Y.~Abdilor, Rock Mech. Rock Eng. \textbf{48}, 843
  (2015)

\bibitem{Pouragha2020}
M.~Pouragha, M.~Eghbalian, R.~Wan, Int. J. Rock Mech. Min. Sci \textbf{125},
  104154 (2020)

\bibitem{Xu2020}
G.~Xu, M.~Gutierrez, C.~He, W.~Meng, Acta Geotech. \textbf{7} (2020)

\bibitem{Jean1999}
M.~Jean, Comput. Methods in Appl. Mech. Eng. \textbf{177}, 235 (1999)

\bibitem{Dubois2018}
F.~Dubois, V.~Acary, M.~Jean, Comptes Rendus - Mecanique \textbf{346}, 247
  (2018)

\bibitem{Nguyen2015}
D.H. Nguyen, E.~Az{\'{e}}ma, P.~Sornay, F.~Radjai, Phys. Rev. E \textbf{91},
  022203 (2015)

\bibitem{Cantor2016}
D.~Cantor, E.~Az{\'{e}}ma, P.~Sornay, F.~Radjai, Comput. Part. Mech.
  \textbf{4}, 441 (2017)

\bibitem{Huillca2020}
Y.~Huillca, M.~Silva, C.~Ovalle, S.~Carrasco, J.~Quezada, G.~Villavicencio,
  Acta Geotech.  (2020)

\bibitem{Jones2019}
D.R. Jones, M.F. Ashby, in \emph{Engineering Materials 1}, edited by D.R.
  Jones, M.F. Ashby (2019)

\bibitem{LMGC90Web}
F.~Dubois, M.~Jean, et~al, \emph{{LMGC90 website}},
  \url{https://git-xen.lmgc.univ-montp2.fr/lmgc90/lmgc90_user} (2020), [Online;
  accessed Nov-2020]

\bibitem{Dubois2011}
F.~Dubois, M.~Jean, M.~Renouf, R.~Mozul, A.~Martin, M.~Bagn{\'{e}}ris,
  \emph{{LMGC90}}, in \emph{10e colloque national en calcul des structures}
  (2011)

\bibitem{Griffith1921}
A.A. Griffith, Philos. Trans. A. Math. Phys. Eng. Sci. \textbf{221}, 163 (1921)

\end{thebibliography}

\end{document}